\documentstyle[11pt]{article}

\begin{document}
\begin{flushright}
CERN-TH.7132/94\\
UND-HEP-94-BIG01\\
TPI-MINN-94/1-T\\
UMN-TH-1234/94\\
TECHNION-PH-94-1\\
January 1994
\end{flushright}
\vspace{0.7cm}
\begin{center}

{\Large \bf NON-LEPTONIC DECAYS OF BEAUTY HADRONS --
FROM PHENOMENOLOGY
TO THEORY
\footnote{To appear in the
second edition of the book `B Decays', S.Stone (ed.), World Scientific}
}
\end{center}
\vspace{0.65cm}
\centerline{\large I. Bigi$^{a,b}$, B. Blok$^c$, M. Shifman$^d$,
N. Uraltsev$^{a,e}$, A. Vainshtein$^{d,f}$}
\vspace{0.4cm}
\centerline{\it $^a$ Theor. Phys. Div., CERN, CH-1211 Geneva 23}
\baselineskip=12pt
\centerline{\it $^b$ Dept. of Phys., Univ. of Notre Dame du Lac,
Notre Dame, IN 46556, USA\footnote{Permanent address}}
\baselineskip=12pt
\centerline{\it $^c$ Phys. Dept., TECHNION, Haifa, Israel}
\baselineskip=12pt
\centerline{\it $^d$ Theor. Phys. Inst., Univ. of Minnesota,
Minneapolis, MN 55455, USA}
\baselineskip=12pt
\centerline{\it $^e$ St. Petersburg Nucl. Phys. Inst., Gatchina,
St. Petersburg 188350, Russia$^2$}
\baselineskip=12pt
\centerline{\it $^f$ Budker Inst. of Nucl. Phys.,
Novosibirsk 630090, Russia}

\vspace{0.4cm}
\centerline{\rm e-mail addresses:}
\baselineskip=12pt
\centerline{\it VXCERN::IBIGI, BIGI@UNDHEP;
PHR34BB@TECHNION.BITNET;}
\baselineskip=12pt
\centerline{\it SHIFMAN@VX.CIS.UMN.EDU;
VAINSHTE@VX.CIS.UMN.EDU;}
\baselineskip=12pt
\centerline{\it
VXCERN::URALTSEV, URALTSEV@LNPI.SPB.SU}

\baselineskip=13pt
\vspace{0.7cm}
\centerline{\Large \bf ABSTRACT}
\vspace*{.5cm}
In the last few years considerable progress has been achieved
in our understanding of the decays of heavy flavour hadrons.
One can now calculate $inclusive$ transition rates in QCD proper
through an expansion in inverse powers of the heavy flavour
quark mass
without recourse to phenomenological assumptions. The
non-perturbative contributions are treated systematically
in this way; they are found to produce corrections of order a
few percent in beauty decays, i.e. typically somewhat smaller than
the perturbative corrections. One finds, among other things:
(a) The lifetime of $B^-$ mesons is
predicted to be longer than that of $B^0$ mesons
by several percent. (b) The QCD  prediction for
the semileptonic branching ratio of $B$
mesons appears to exceed present experimental values.
We discuss the implications of this discrepancy.
The phenomenological engineering that has been developed for the
description of {\em exclusive} two-body modes of $B$ mesons has
reached a mature stage and awaits more precise and detailed
experimental tests. First steps towards a genuine QCD treatment
of these modes are being made.

\large

\def\lsim{\mathrel{\rlap{\lower3pt\hbox{\hskip0pt$\sim$}}
    \raise1pt\hbox{$<$}}}         %less than or approx. symbol
\def\gsim{\mathrel{\rlap{\lower4pt\hbox{\hskip1pt$\sim$}}
    \raise1pt\hbox{$>$}}}         %greater than or approx. symbol
\def\dblint{\mathop{\rlap{\hbox{$\displaystyle\!\int\!\!\!\!\!\int$}
}
    \hbox{$\bigcirc$}}}
\def\ut#1{$\underline{\smash{\vphantom{y}\hbox{#1}}}$}

\newcommand{\ind}[1]{_{\begin{small}\mbox{#1}\end{small}}}
\newcommand{\WA}{{\em WA}}
\newcommand{\SM}{Standard Model }
\newcommand{\QCD}{{\em QCD }}
\newcommand{\KM}{{\em KM }}
\newcommand{\hscale}{\mu\ind{hadr}}
\newcommand{\aver}[1]{\langle #1\rangle}

\newcommand{\appa}{\mbox{\ae}}
\newcommand{\CP}{{\em CP}}
\newcommand{\fy}{\varphi}
\newcommand{\hi}{\chi}
\newcommand{\al}{\alpha}
\newcommand{\as}{\alpha_s}
\newcommand{\gf}{\gamma_5}
\newcommand{\gm}{\gamma_\mu}
\newcommand{\gn}{\gamma_\nu}
\newcommand{\be}{\beta}
\newcommand{\ga}{\gamma}
\newcommand{\de}{\delta}
\renewcommand{\Im}{\mbox{Im}}
\renewcommand{\Re}{\mbox{Re}}
\newcommand{\GeV}{\,\mbox{GeV }}
\newcommand{\MeV}{\,\mbox{MeV }}
\newcommand{\matel}[3]{\langle #1|#2|#3\rangle}
\newcommand{\state}[1]{|#1\rangle}
\newcommand{\ra}{\rightarrow}
\newcommand{\ve}[1]{\vec{\bf #1}}

\large
\addtocounter{footnote}{-2}

\vspace*{.7cm}
\begin{flushright}
{\em `Anyone who keeps the ability to see beauty never grows old'}
\hspace*{1em} Franz Kafka
\end{flushright}

%\vspace*{.4cm}
\newpage

\section{Introduction}

In the two years since the appearance of the first edition of this book
our understanding of the decays of beauty hadrons has been
improved
quite significantly. The availability of better and more
comprehensive
data has obviously helped here -- but so has the emergence of more
reliable theoretical tools! Rather than having to rely on
phenomenological models we can now employ approaches that are
directly based on QCD with no appeal to a `deus ex machina', at least
for semileptonic and inclusive non-leptonic decays. Therefore the
article on non-leptonic beauty decays had to be re-written
completely. {\em Inclusive} decays of beauty hadrons can now be
treated in QCD in a quantitative fashion with systematic estimates
of the uncertainties.

The employed formalism combines the heavy quark mass
expansion with other elements derived from QCD proper. Its
technical foundation is the Wilson operator product expansion
(OPE)~\cite{wilson}. The idea of applying OPE to the inclusive heavy
flavour decays had emerged in
in the eighties~\cite{bilic} and has now grown into a
well-developed scheme~\cite{chay,bigi0,bigi1,bigi2}.

The basic procedure can be illustrated by a simple analogy
with nuclear $\beta$ decay. There are two
effects distinguishing leptons in the decay of a neutron bound
inside a nucleus from those in a decay of a free neutron:

(a) nuclear binding effects;

(b) Pauli statistics correlating the electrons surrounding
the nucleus with those emerging from the $\beta$
decay.

The typical energies of the bound electrons $\epsilon_{el}$
are certainly small
compared to $E_{rel}$, the energy released in the decay;
let us assume -- although this is not true in
reality -- that also the nuclear binding energies $\epsilon_{nucl}$
were small compared to $E_{rel}$. In that case nuclear $\beta$
decays would proceed like the decays of {\em free}
neutrons to a good approximation; corrections to this
simple `spectator' picture could be computed via an expansion in
powers of $\epsilon_{nucl}/E_{rel}$, $\epsilon_{el}/E_{rel}$
In practice, however, the corrections for nuclear $\beta$ decay
are incorporated by explicitly
using the wavefunctions of the bound nucleons and
electrons.

There arise analogous corrections to the decay rate
for a quark $Q$ inside a hadron:

(a) interactions of the decaying quark with other partons in
the hadron\footnote{There is also the annihilation of the
heavy quark with the light (spectator) antiquark for which an analogy
is found in the K capture of bound electrons by a heavy nucleus.}.

(b) Pauli interference effects of the decay products with
other partons in the hadron; e.g.: $b\bar u\ra c\bar u d \bar u$.

The difference with the example above is quite obvious:
even in the limit $m_Q\ra\infty$ a non-relativistic bound state
treatment is inapplicable since the dynamical degrees of freedom of
the heavy flavour hadron $H_Q$ cannot be fully
described by a hadronic
wavefunction! We will return to this point later on.
The
most reliable approach is then to evaluate weak decay rates
of heavy flavour hadrons through an expansion in
powers of $\mu _{had}/m_Q$; $m_Q$ -- the heavy flavour quark
mass --
is a measure of the energy release in the decay and
$\mu _{had}$ represents ordinary hadronic scales which enter
through the matrix elements for heavy flavour hadrons. Since
$\mu _{had}< 1\; \GeV$ (details will be given below) one expects
such an expansion to work quite well for beauty decays.

The example given above illustrates two important features of our
analysis: it applies to {\em inclusive} non-leptonic and
semileptonic transitions, and the usefulness of such an expansion
rests on a large {\em energy release} in the decay; i.e. in
$Q_1\ra Q_2`W$', where  $`W$' denotes $q^{\prime}\bar q$ or
$l\nu$ one requires
$m_{Q_1}- m_{Q_2}\gg \mu _{had}$. This is in contrast to the
approach usually referred to as the heavy quark
symmetries~\cite{HQS}
 which applies
to {\em exclusive} semileptonic amplitudes provided that
the initial and the final  quarks are both  heavy,
i.e. $m_{Q_1}, \; m_{Q_2}\gg \mu _{had}$ but the ratio $m_{Q_2}/
m_{Q_1}$ is otherwise arbitrary.  The QCD-based description of the
inclusive decays of heavy flavours naturally incorporates many
elements of Heavy Quark Effective Theory
({\em HQET})~\cite{EHQT,EHQTr}. The systematic expansion in $m_Q^{-1}$ is
inherent to both, and it
allows us -- as we will see shortly -- to determine the size
of important
input parameters for our analysis.

This survey will be organized as follows: in Sect. 2 we review
the general features of the $1/m_Q$ expansion for inclusive
decay rates; in Sect. 3 we present quantitative predictions for
lifetime ratios, semileptonic branching ratios and radiative widths
and discuss their theoretical uncertainties; in Sect. 4 we address
exclusive non-leptonic decays before giving a summary and an
outlook
in Sect. 5.

\section{Inclusive Decay Rates of Beauty Hadrons: Formalism}

As mentioned above, the systematic approaches to inclusive
heavy flavor decays date back to the beginning of the eighties.
Unitarity relates inclusive decay rates
to the imaginary part of certain forward `scattering'
amplitudes; this is a rather trivial observation~\cite{SV1},
yet it
opened the way for the consistent use of OPE.  The starting object of
our analysis is the
transition operator $\hat T(b\ra f\ra b)$ describing the forward
scattering of $b$ quarks via an intermediate state $f$.  To second
order
in the weak interactions the transition operator is given
by~\cite{SV1}
$$
\hat T(b\ra f\ra b)= i\int d^4x\{ {\cal L}_W(x)
{\cal L}_W^\dagger (0)\}_T ,
\eqno(1)$$
where ${\cal L}_W$ denotes the relevant effective weak
Lagrangian and $\{.\}_T$ is the time-ordered product.
Treating $m_b$ as a large parameter a Wilson
OPE allows expressing the non-local
operator $\hat T$ as an infinite sum of local operators of increasing
dimension with coefficients containing higher and higher powers of
$1/m_b$. The lowest dimensional term will dominate in the limit
$m_b\ra \infty$; for beauty decays
that is the dimension three operator $\bar bb$.
The width for
the decay of a beauty hadron $H_b$ into an inclusive
final state $f$ is obtained by taking the expectation value of
$\hat T$ between the state $H_b$. Through order $1/m_b^3$ one
finds~\cite{bilic,bigi0,bigi1,bigi3}:
$$\Gamma (H_b\ra f)=\frac{G_F^2m_b^5}{192\pi ^3}|KM|^2
\left[ c_3(f)\frac{\matel{H_b}{\bar bb}{H_b}}{2M_{H_b}}+
\frac{c_5(f)}{m_b^2}\frac{
\matel{H_b}{\bar bi\sigma _{\mu \nu}G_{\mu
\nu}b}{H_b}}{2M_{H_b}}+ \right.
$$
$$\left. +\sum _i \frac{c_6^{(i)}(f)}{m_b^3}\frac{\matel{H_b}
{(\bar b\Gamma _iq)(\bar q\Gamma _ib)}{H_b}}
{2M_{H_b}} + {\cal O}(1/m_b^4)\right] ,  \eqno(2)$$
where the dimensionless coefficients $c_i(f)$ depend on the
parton level characteristics of $f$ (such as the ratios of the
final state quark masses to $m_b$), $KM$ denotes the
appropriate combination of weak mixing angles and
$G_{\mu \nu}$ the gluonic field strength tensor.
The last term implies also the summation over the four-fermion
operators with different light flavours $q$.
Notice that
the factor $1/2M_{H_b}$ reflects the relativistic normalization of
the state $|H_b\rangle$.

It is through the expectation values of local operators
appearing on the right-hand side of eq. (2)
that the dependence on the decaying $hadron$, and on
non-perturbative forces in general, enters instead
of entering through the  wavefunctions as in nuclear $\beta$ decay.
Since these are matrix elements for real $b$ hadrons one
sees that $\Gamma (H_b\ra f)$ is indeed expanded into
a power series in $\mu _{had}/m_b$. The heavy quark expansion
enables us to express decay rates in terms of
 the expectation values of local operators
taken between beauty hadrons. Using heavy quark expansions
one can relate
some of these hadronic expectation values
 -- and in particular those that appear
in the leading terms of the expansion in eq. (2) -- to
other observables and thus extract their size, as we will discuss
now.

\noindent (i) To determine $\matel{H_b}{\bar bb}{H_b}$ one
expresses, {\em via} equations of motion, the
operator $\bar bb$ through another series in inverse powers
of $m_b$ (which constitutes a static expansion)~\cite{bigi1}:
$$
\langle H_b|\bar b b|H_b \rangle =
$$
$$
\langle H_b| v_{\mu}\bar b\gamma _{\mu}b-
\frac{1}{2m_b^2}\bar b[(iv\cdot D)^2-(iD)^2]b +
\frac{1}{4m_b^2}\bar bi\sigma \cdot Gb|H_b \rangle +
{\cal O}(1/m_b^4)
\;\; \eqno (3)
$$
with $v_{\mu}$ denoting the   four-velocity of the heavy hadron.
The first operator appearing on the $RHS$ of eq. (3) is
actually the Noether current for the (global) heavy
flavour quantum
number; its expectation value is thus determined by the beauty
content of $H_b$ and therefore
$$
\matel{H_b}{\bar bb}{H_b}/(2M_{H_b})= 1 + {\cal O}(1/m_b^2).
\eqno(4)$$
It is
this term that exactly reproduces the parton model spectator
result in the limit $m_b\ra \infty$, which attributes equal
lifetimes to all hadrons of a given heavy flavour.

\noindent (ii) The chromomagnetic operator
$\bar bi\sigma \cdot Gb$ appears directly in eq. (2) and
indirectly through the expansion of $\bar bb$. Its expectation
value vanishes for the baryon $\Lambda _b$:
$$ \matel{\Lambda _b}{\bar bi\sigma \cdot Gb}{\Lambda _b}
\simeq 0\eqno(5)$$
For the $B$ meson it is given by the hyperfine splitting of
the $B^*$ and $B$ masses:
$$
\matel{B}{\bar bi\sigma \cdot Gb}{B}/(2M_{H_b})\simeq
\frac{3}{2}(M^2_{B^*}-M^2_B)\simeq 0.74\,\GeV^2
\eqno(6)$$

\noindent (iii) The second operator on the $RHS$ of eq. (3)
describes the kinetic energy of the $b$ quark moving under the
influence of the non-trivial gluon background field prevalent inside
the hadron $H_b$:
$\matel{H_b}{\bar b[(iv\cdot D)^2-(iD)^2]b}{H_b}\simeq
\matel{H_b}{\bar b(i\vec D)^2b}{H_b}\equiv
\langle (\vec p_b)^2\rangle _{H_b}\cdot (2M_{H_b})$. Its appearance in eq.(3)
has a
very transparent meaning:
$$
\frac{\matel{H_b}{\bar bb}{H_b}}{2M_{H_b}}=1-
\frac {\langle (\vec p_b)^2\rangle _{H_b}}{2m_b^2}+ \frac {3}{8}
\frac {M_{B^*}^2-M_B^2}{m_b^2}+{\cal O}(1/m_b^3)
\eqno(7)$$
The first two terms on the $RHS$ of eq. (7) represent the mean value
of the  factor $\sqrt{1-\vec v^2}$ due to time dilation slowing down
the decay of the $b$ quark in a moving frame. The
heavy quark expansion relates the
{\em difference} in the expectation values of the kinetic energy
operator for heavy flavour baryons and mesons to the masses of the
charm and beauty baryons and mesons~\cite{buvpre}:
$$\langle (\vec p_b)^2\rangle _{\Lambda
_b}-
\langle (\vec p_b)^2\rangle _B\simeq
\frac{2m_bm_c}{m_b-m_c}\cdot \{ [\langle M_B\rangle -
M_{\Lambda _b}]-[\langle M_D\rangle -
M_{\Lambda _c}]\} \eqno(8)$$
where $\langle M_{B,D}\rangle$ denote the `spin averaged' meson
masses,
$$
\langle M_B\rangle \equiv \frac{1}{4} (M_B+3M_{B^*})
\eqno(9)$$
and likewise for $\langle M_D\rangle$. Eq. (9) implies that
$c$ quark can be also treated as heavy so that
$\langle (\vec p_b)^2\rangle_{H_b}=
\langle (\vec p_c)^2\rangle_{H_c}$.
{}From the present data we
obtain
$$\langle (\vec p_b)^2\rangle _{\Lambda_b}-
\langle (\vec p_b)^2\rangle _B\simeq
(-0.03\pm 0.17)\; \GeV^2
\eqno(10)$$
where the error is due to the $\pm 30\; \MeV$ experimental
uncertainty in $M_{\Lambda _b}$. The $\Lambda _b$ mass has to be
measured with better than a 10 MeV precision to make this relation
numerically useful.

Using the commutator algebra of the covariant derivatives
$iD_{\mu}$ one
can derive an `uncertainty principle' for their components and thus
establish a model independent lower bound~\cite{motion}:
$$
{\langle (\vec p_b)^2\rangle _B}
\geq 0.18\; (\GeV )^2
.\eqno(11)$$
An existing
analysis based on QCD sum rules yields a value only three times
larger than this lower limit~\cite{braun}:
$$
\langle (\vec p_b)^2\rangle _B
\simeq 0.6\; (\GeV )^2.
\eqno(12)$$
(iv) The expectation value for the four-quark operators looks
very similar
to the one controlling $B^0-\bar B^0$ oscillations:
$$
\matel{B(p)}{(\bar b_L\gamma _{\mu}q_L)(\bar q_L\gamma
_{\nu}b_L)}
{B(p)}\simeq f_B^2 p_{\mu}p_{\nu}
\eqno(13)$$
where we have set the so-called bag factor to unity.

Before discussing the phenomenology that is obtained from eq.(2) we
want
to point out seven basic observations:

(1) The most important aspect -- quantitatively as well as
conceptually -- of the expression in eq. (2) is contained in the
element that is missing there: there are no non-perturbative
contributions
of order $1/m_b$ to fully integrated rates~\cite{bigi1}!
The numerical impact of this
fact is obvious: since the leading non-perturbative corrections then
arise on the $1/m_b^2$ level, they fade away quickly with increasing
heavy flavour quark mass and amount, for the beauty decays,
 to effects of order of
several
percent only since  the scale is set by the
quantities
$$G_B\equiv \frac{\matel{B}{\bar bi\sigma \cdot Gb}{B}}
{2m_b^2\cdot (2M_{H_b})}\simeq 0.015,
\eqno(14a)$$
$$
K_B\equiv \frac{\langle \vec p_b^2\rangle _B}
{m_b^2}
\sim 0.015.
\eqno(14b)$$
Let us note in passing that the analogous contributions in charm
decays are much larger since they are amplified by a factor
$(m_b/m_c)^2\sim 10$!

%\noindent
The conceptual relevance of the absence of $1/m_b$ terms is
of a more subtle, but not less important nature. On the one hand it
confirms the conjecture that the (current) quark mass $m_b$ rather
than the hadron mass $M_B$ represents the natural expansion
parameter.
For if the total width were correctly expressed in terms of $M_B$
then
a term linear in $1/m_b$ had to appear: $\Gamma (B) \propto
G_F^2M_B^5=G_F^2m_b^5(1+5\bar \Lambda _B /m_b+...)$ with the
notation $M_B=m_b+\bar \Lambda _B +{\cal O}(1/m_b)$.
This is actually more than an academic point for it would
have a significant impact on the lifetime difference
between $B$ mesons and $\Lambda _b$ baryons:
$(\tau (\Lambda _b)-\tau (B))/\tau (B) \propto 1/m_b$ since
${\bar\Lambda}_{\Lambda_b}\neq \bar \Lambda _B$!
On the other
hand it is
quite instructive to understand the dynamical reason why no linear
$1/m_b$ terms arise in total decay rates whereas they do appear in
mass formulae. Looking at the explicit QCD calculation given in
ref.~\cite{bigi1}
one can already infer why no contribution of order $1/m_b$ arises:
only a dimension four operator could generate such a term and there
simply does not exist such an appropriate operator that is gauge
invariant and cannot be absorbed into the dimension three operator
$\bar bb$ by means of equations of motion.
(This means that proper care has to be applied in
employing
a consistent definition of the heavy flavour mass $m_b$.) It is
therefore
the colour symmetry, i.e. the fact that the colour flow is
conserved
that ensures the absence of a $1/m_b$ correction.
This correction is absent also for differential distributions like
the lepton spectra in semileptonic decays {\it outside} the
end-point region. The size of this end-point domain is of order
$\mu_{hadr}/m_b$ if all energies are measured in units
of $m_b$. In this domain we cannot limit ourselves to the operators
of the lowest dimension, and one needs to sum up an infinite series to
determine the shape of the distribution which  leads to
modification of the spectrum of order  unity in the end-point
region. It is remarkable that
the full integral over the
end-point region still has no corrections proportional
to $1/m_b$!

(2) The distinction between beauty baryon and meson decays is
systematically expressed through differences in the appropriate
expectation values of the same operators, see eq.(2). Differences
arise first in the leading non-perturbative corrections of order
$1/m_b^2$, see eqs.(5, 6, 8). Yet apart from some small $SU(3)_{FL}$
breaking effects they affect $B_d$, $B^-$ and $B_s$ decays in the
same way; likewise for $\Lambda _c$ vs. $D^0$, $D^+$ and $D_s$
decays. Numerically they are comparable to -- actually typically
smaller than -- $perturbative$ corrections in beauty decays whereas
they dominate perturbative effects in charm decays.

(3) Local four-quark operators of dimension six finally produce
differences between all the $B$ meson lifetimes:
$$\frac{\Gamma _{nonspect}(B)}{\Gamma _{spect}(B)}\propto
\frac{f_B^2}{m_b^2}$$
which formally scales like $1/m_b^3$. These effects are therefore
predicted to be greatly
reduced relative to the considerable lifetime differences in the
$D$ system.

(4) Contributions of order $1/m_b^4$ are generated by dimension
seven operators. Yet it appears to be practically unfeasible to
determine the expectation values for all or even most of them. What
seems possible -- although it has not been done yet -- is to analyze
a small subset of them, namely those yielding $factorizable$
contributions, for a more detailed error estimate.

(5) It is intuitively obvious that the $b$ quark does not rest inside a
beauty hadron $H_b$, but will move around with a certain `Fermi
momentum'.
This has been incorporated into phenomenological models of
inclusive
heavy flavour decays, first in ref.~\cite{alta}.
It has been first stated in ref.~\cite{bigi2}  and then further discussed in
refs.~\cite{mn,motion,neubert,wise} that this notion
of Fermi motion of a heavy flavour quark finds a natural home also
in a rigorous QCD treatment; yet, strictly speaking, no hadronic
wavefunction for $H_b$ in the usual sense
can be found that reproduces the Fermi
motion
beyond the second moment $\aver{(\vec p)^2}$. We will not
discuss
this in any detail here since $\aver{(\vec p)^2}$ has a reduced
numerical relevance in fully integrated rates (although it is of crucial
importance for shaping the end-point spectrum in the inclusive
semileptonic and radiative decays).

(6) While our expressions for inclusive decay rates are firmly based
on
QCD, one has to invoke explicitly -- and at the moment additionally --
the concept of duality. The operator product expansion is unambigously
defined in the Euclidean domain, yet the kinematics of
the
actual decay are of a time-like Minkowskian nature. It is then
conceivable -- although it has never been illustrated in a
clear way -- that a
translation between the Euclidean and the Minkowskian expression
that is based on $local$ duality
does
not hold in non-leptonic or for that matter even in semileptonic
decays. We view this as a mathematical rather than as a physical
caveat. The conjecture of duality can be supported by some general
arguments,
yet their discussion would lead beyond the scope of this article.

(7) The expression in eq. (2) is based on OPE where one separates
short
distance and long distance dynamics by isolating the latter in the
local
operators and their matrix elements and letting the former
determine
the c number coefficients $c_i(f)$. In actual calculations one goes one
step further: one computes the coefficients $c_i(f)$ in perturbation
theory alone although non-perturbative short distance contributions do
exist. The latter  are guestimated (and in $\tau$ decays found~\cite{pich})
to be
quite small. We adopt this procedure which we refer to as the
`Standard Version' of OPE although we will also comment on possible
limitations later on.

\section{Phenomenology of Inclusive Beauty Decays}

There are five types of inclusive observables we will discuss here,
namely: (1) total lifetimes; (2) semileptonic branching ratios;
(3) other inclusive non-leptonic decays;
(4) radiative decays; (5) charm multiplicity in the
final state and (6) $B\ra$ charmonia + X.

\subsection{Total Lifetimes}

There exists a triple motivation behind measuring the lifetimes
of different species of beauty hadrons as
precisely as possible: (a) Representing the most inclusive quantity it
provides a clear and well-defined challenge to theory. (b) It allows
to obtain the semileptonic {\em width} from the measured
semileptonic branching ratios; from this width one extracts
the KM parameter $|V_{cb}|$ etc. (c) It is a pre-requisite of a
detailed analysis of $B^0-\bar B^0$ oscillations.

There is no basic uncertainty about the weak forces driving
non-leptonic beauty decays: at the scale $M_W$ they are given
by the Lagrangian
$${\cal L}^{\Delta B=1}_W(\mu = M_W)=\frac{4G_F}{\sqrt{2}}
[V_{cb}\bar c_L\gamma _{\mu}b_L+V_{ub}\bar u_L\gamma
_{\mu}b_L]
\cdot [V^*_{ud}\bar d_L\gamma _{\mu}u_L+V^*_{cs}\bar s_L\gamma
_{\mu}
c_L] .
\eqno(15)$$
where we have ignored Cabibbo suppressed transitions
and also the $b\ra t$ coupling
since we will not discuss $B^0-\bar B^0$ oscillations and Penguin
contributions here.
Radiative QCD corrections lead to a well-known renormalization at
scale
$m_b$, which is often referred to
as ultra-violet (UV) renormalization:
$${\cal L}^{\Delta B=1}_W(\mu =
m_b)=\frac{4G_F}{\sqrt{2}}V_{cb}V^*_{ud}
\{ c_1(\bar c_L\gamma _{\mu}b_L)(d_L\gamma _{\mu}u_L)+
c_2(\bar d_L\gamma _{\mu}b_L)(\bar c_L\gamma
_{\mu}u_L)\}\eqno(16)$$
for $b\ra c\bar ud$ and likewise for $b\ra c\bar cs$ etc.
transitions; the
QCD corrections are lumped together into the coefficients $c_1$ and
$c_2$ with
$$c_1=\frac{1}{2}(c_++c_-), \; \; c_2=\frac{1}{2}(c_+-c_-)\eqno(17a)$$
$$c_{\pm}=[\frac{\alpha _S(M_W^2)}{\alpha _S(m_b^2)}]^{\gamma
_{\pm}},
\; \; \gamma _+=\frac{6}{33-2N_f}=-\frac{1}{2}\gamma _-
\eqno(17b)$$
in the leading log approximation. Numerically this amounts to
$$c_1(LL)\simeq 1.1,\; \; c_2(LL)\simeq -0.23\eqno(18)$$
Including next-to-leading log corrections one obtains
$$c_1(LL+NLL)\simeq 1.13,\; \;
c_2(LL+NLL)\simeq -0.29\eqno(19)$$
i.e. a mild enhancement of the original  coupling
together with the appearence of
 an induced  operator with a different color flow.
Later we will also include~\cite{hybrid} the so-called `hybrid' renormalization
reflecting radiative corrections in the domain from $m_b$ down to
$\mu _{had}$~\cite{hybrid1}.

As already stated in Sect. 2 differences
between $B$ $meson$ lifetimes arise on the $1/m_b^3$ level
generated by local four-quark operators $(\bar b_L\gamma
_{\mu}q_L)
(\bar q_L\gamma _{\nu}b_L)$. Based on this scaling law one can
already
infer from the observed $D$ meson lifetime ratios that the various
$B$
meson lifetimes will differ by no more than 10 percent or so.

In phenomenological models two distinct mechanisms producing
lifetime differences had been noted,
namely

$\bullet$ Weak Annihilation (WA) and

$\bullet$ Pauli Interference (PI)~\cite{pi}

\noindent
in qualitative analogy to the situation in nuclear $\beta$ decay
as explained in the Introduction.
In the $1/m_b$ expansion they emerge as follows. There are two
types of four-quark operators which are distinguished by how the
light quark flavours are connected inside the hadron $H_b$. This can
be seen from Figs.~1. Upon
integrating out the $c$, $\bar q$ and $q'$ fields
in the diagram of Fig. 1a  where the square boxes represent
${\cal L}_W^{\Delta B=1}$ one obtains the operator $\bar bb$; cutting then
the $q'$ line in Fig. 1a
and
connecting it to the $q'$ constituent of the $B$ meson, as shown in
Fig. 1b, one has a WA transition operator; cutting instead the $\bar q$ line
and
connecting it to the $B$ constituents, see Fig. 1c, leads to
the four-fermion operator describing PI.

It turns out that the WA processes can change
$B$ lifetimes by no more than, say, 1\%; due to interference with the
spectator reaction they could actually {\em prolong} $\tau (B_d)$
relative to $\tau (B^-)$ rather than reduce it~\cite{bigi0}!
The dominant effect is
provided by PI which produces an additional contribution to the
$B^-$ width:
$$\Gamma (B^-)=\Gamma _{Spect}(B)+\Delta \Gamma _{PI}(B^-
)\eqno(20a)$$
$$\Delta \Gamma _{PI}(B^-)\simeq \Gamma _0\cdot
24\pi ^2 \frac{f_B^2}{M_B^2}[c_+^2-c_-^2+\frac{1}{N_C}(c_+^2+c_-^2)],\;
\; \Gamma _0=\frac{G_F^2m_b^5}{192\pi ^3}|V(cb)|^2\eqno(20b)$$
where the appearance of $f_B$ reflects the fact that PI -- like WA --
requires the spatial overlap of two (anti)quark fields.

Eq. (20b) exhibits an intriguing result: $\Delta \Gamma _{PI}(B^-)$
is positive for $c_+=1=c_-$, i.e. PI acts
$constructively$. Radiative QCD corrections with
$c_+\simeq 0.84$, $c_-\simeq 1.42$ turn PI into a
$destructive$ interference which prolongs $\tau (B^-)$ by a
tiny amount. In eq. (20b)
only UV renormalization has been applied. Hybrid
renormalization amplifies this effect considerably
and one obtains~\cite{bigi0,hybrid}:
$$\Delta \Gamma _{PI}(B^-)\simeq \Gamma _0\cdot
24\pi ^2 \frac{f_B^2}{M_B^2}
\kappa ^{-4}[(c_+^2-c_-^2)\kappa ^{9/2}+\frac{c_+^2+c_-^2}{3}-
\frac{1}{9}(\kappa ^{9/2}-1)(c_+^2-c_-^2)], $$
$$\kappa \equiv [\frac{\alpha _S(\mu ^2_{had})}{\alpha
_S(m_b^2)}]^{1/b},\;
b=11-\frac{2}{3}n_F\eqno(21)$$
Altogether one finds:
$$\frac{\tau (B^-)}{\tau (B_d)}\simeq
1+0.05\cdot \frac{f_B^2}{(200\; \MeV )^2}\; ,\eqno(22)$$
i.e. the lifetime of a charged $B$ meson is predicted to exceed that of
a neutral $B$ meson by several percent, but not more
than
ten percent~\footnote{It should be noted that an analogous analysis
yields
$\tau (D^+)/\tau (D^0)\sim 2$, which is quite consistent with the
observed value of $\simeq 2.5$. Yet one has to keep
in mind that the charm
quark mass is not much larger than typical hadronic masses; the
convergence of the $1/m_c$ expansion is thus too slow, if it happens at
all, to yield a better than semi-quantitative description of non-leptonic
charm decays. For a recent review of
the theoretical situation here see ref.~\cite{bloks}. }. Corrections of
order $1/m_b^4$ which have been ignored here
are unlikely to change this prediction significantly;
this statement will however be qualified below in our
discussion of the semileptonic $B$ branching.

One also expects
$$\bar \tau (B_d)\simeq \bar \tau (B_s)\eqno(23)$$
where $\bar \tau$ denotes the average lifetime of the two mass eigenstates in
the
$B^0-\bar B^0$ system.
It is at least amusing to note that the largest lifetime
difference among $B$ mesons is most likely
produced by a
subtle mechanism, namely $B_s$-$\bar B_s$ oscillations
with both PI-like and WA-like mechanisms contributing~\cite{bsosc}:
$$\frac{\Delta \Gamma (B_s)}{\bar \Gamma (B_s)}
\equiv \frac{\Gamma (B_{s,short})-\Gamma (B_{s,long})}
{\bar \Gamma (B_s)}\simeq 0.18\cdot
\frac{f_{B_s}^2}{(200\; \MeV )^2}\; .\eqno(24)$$
One can search for the existence of two
different $B_s$ lifetimes by
comparing $\tau (B_s)$ as measured in
$B_s\ra \psi \phi$ and in $B_s\ra l\nu X$.
Analogously one can compare
$\tau (B_d)$ as obtained from $B_d\ra \psi K_S$, from
$B_d\ra \psi K^*$ and from $B_d\ra l\nu X$. In the
$B_d$ case
one theoretically
expects a lifetime difference on the percent level only.
Whether an
effect of the size indicated in eq. (24) is large
enough to be ever observed in a real experiment is of
course a different matter. It has to be said, though,
that eq.(24) does not represent a `gold-plated' prediction.
It is conceivable that the underlying computation
underestimates the actual
lifetime difference.

No detailed analysis has been performed yet on $\tau (\Lambda
_b)$;
simple estimates lead to the expectation
$$\frac{\tau (\Lambda _b)}{\tau (B_d)}
\sim 0.9 .\eqno(25)$$

Present measurements yield~\cite{sharma}
$$\tau (B^+)=1.66\pm 0.11\;  psec\eqno(26a)$$
$$\tau (B_d)=1.51\pm 0.10\;  psec\eqno(26b)$$
$$\frac{\tau (B^+)}{\tau (B_d)}=1.12\pm 0.09\eqno(26c)$$
$$\tau (B_s)=1.54\pm 0.24\;  psec\eqno(26d)$$
$$\tau (\Lambda _b)=1.07\pm 0.16\;  psec\eqno(26d)$$
While the predictions stated above on the lifetime
ratios for $B$ mesons, see eqs.(22,23),
are quite consistent with the measurements, one cannot draw a
definite conclusion at the moment. The $\Lambda _b$ lifetime
appears to be shorter than the $B$ lifetimes, though:
$$\tau (B^+)-\tau (\Lambda _b)=0.59\pm 0.19\; psec\eqno(27)$$
$$\tau (B_d)-\tau (\Lambda _b)=0.44\pm 0.19\; psec\eqno(28)$$
Qualitatively this is expected
although the size of the difference seems to be
larger than anticipated.

\subsection{Semileptonic Decays}

The semileptonic branching ratio of beauty quarks depends on
fundamental quantities of the Standard Model, namely KM
parameters and quark masses. It is then the primary goal
of the measurements to
determine the numerical values of these quantities from the
data. The theoretical challenge on the other hand consists in
disentangling the effects of the strong interactions both in
their perturbative as well as non-perturbative aspects.

The recent experimental studies have reached a new level of
accuracy and reliability: a `model-independent' ARGUS analysis
yields~\cite{argus}
$$BR_{SL}(B)=9.6\pm 0.5\pm 0.4 \%\eqno(29a)$$
whereas the CLEO collaboration finds~\cite{cleo}
$$BR_{SL}(B)=10.65\pm 0.05\pm 0.33 \%\eqno(29b)$$
using the model of Altarelli et al. for the $shape$ of the lepton
spectrum.
It is an intrinsically phenomenological description, yet
one should keep in mind that it provides a practically
good approximation to the true QCD lepton spectrum as calculated
through
a $1/m_b$ expansion~\cite{bigi2}. The present data thus clearly
suggest:
$$BR_{SL}(B)|_{exp}\leq 11\%\eqno(30)$$
In a naive parton model where even perturbative QCD is ignored one
obtains
$$BR_{SL}(b\ra cl\nu)\simeq 15\div 16\% \eqno(31)$$
i.e. a non-leptonic enhancement of $\sim 50\%$ has to be found to
reproduce the data.
At first sight this would not seem to represent a stiff challenge --
yet so far we have failed to meet it!

There are non-perturbative as well as perturbative corrections
to the semileptonic beauty branching ratio; our ignorance about
the former was very considerable before the arrival of the
Heavy Quark Expansion; we will discuss them
in sequence.

\subsubsection{Non-perturbative Corrections to $BR_{SL}$}

The semileptonic and non-leptonic widths through order $1/m_b^2$
are given by~\cite{bigi1,puzzle}:
$$\Gamma _{SL}(B)=\Gamma _0\cdot
\frac{\matel{B}{\bar bb}{B}}{2M_B}\cdot
\left[ I_0(x,0,0)+\frac{\mu _G^2}{m_b^2}(x\frac{d}{dx}-2)I_0(x,0,0)
\right] ,
\eqno(32)$$
$$\Gamma _{NL}(B)=\Gamma _0\cdot N_C\cdot
\frac{\matel{B}{\bar bb}{B}}{2M_B}\cdot
\left\{ A_0[\Sigma I_0(x)+\frac{\mu _G^2}{m_b^2}(x\frac{d}{dx}-2)
\Sigma I_0(x)]-\right.
$$
$$\left. 8A_2\frac{\mu _G^2}{m_b^2}\cdot
[I_2(x,0,0)+I_2(x,x,0)]\right\} .
\eqno(33)$$
where the following notations have been used:
$I_0$ and $I_2$ are phase-space factors:
$$I_0(x,0,0)=(1-x^2)(1-8x+x^2)-12x^2\ln x,\;
I_2(x,0,0)=(1-x)^3,\; x=(m_c/m_b)^2\eqno(34)$$
$$I_0(x,x,0)=v(1-14x-2x^2-12x^3)+24x^2(1-x^2)\ln
\frac{1+v}{1-v},\; v=\sqrt{1-4x}\eqno(35a)$$
$$I_2(x,x,0)=v(1+\frac{x}{2}+3x^2)-3x(1-2x^2)\ln
\frac{1+v}{1-v}\eqno(35b)$$
with $I_{0,2}(x,x,0)$ describing the $b\ra c\bar c s$ transition,
and $\Sigma I_0(x)\equiv I_0(x,0,0)+I_0(x,x,0)$;
$A_0=\eta J$, $A_2=(c_+^2-c_-^2)/6$ where $\eta =(c_-^2+2c_+^2)/3$,
and
$J$ represents the effect of the subleading
logarithms~\cite{petrarca} (unknown for $A_2$). Moreover,
$$\mu _G^2\equiv \frac{1}{2M_B}
\matel{B}{\frac{1}{2}\bar bi\sigma \cdot Gb}{B}.
\eqno(36)$$

The matrix element $\matel{B}{\bar bb}{B}$
enters as an overall factor into both the semileptonic
and non-leptonic width; its value does therefore not affect the
semileptonic branching ratio. On the other hand
$\matel{B}{\bar bi\sigma \cdot Gb}{B}$ which is determined from
the
observed $B^*-B$ mass splitting does, and it actually reduces
$BR_{SL}(B)$ since $A_2 <0$! Using
$$m_b^{pole}=4.8\; \GeV , \; m_c^{eff}\sim 1.35\; \GeV\eqno(37)$$
one finds
$$\delta BR_{SL}(B)|_{non-pert}\sim -0.02\cdot BR_{SL}(B)\simeq
-0.003\; ,\eqno(38)$$
i.e. a very small reduction!~\footnote{It should be noted that
the corresponding effect is much larger in $D$ decays -- it leads to
a roughly 50 \% reduction in $BR_{SL}(D)$ and is essential for a
self-consistent understanding of charm decays.}
Corrections of order $1/m_b^3$ have been analyzed in
ref.~\cite{puzzle};
as expected, they are estimated to be quite insignificant.
This means that non-perturbative corrections can to a good
first approximation be ignored in $BR_{SL}(B)\;$!

\subsubsection{Perturbative Corrections to $BR_{SL}$}

The preceding discussion shows that it is mainly the
$perturbative$ corrections that control the size of $BR_{SL}(B)$.
They indeed generate a sizeable non-leptonic enhancement thus
reducing $BR_{SL}(B)$, as desired; yet numerically they fall
short of the goal. For one finds~\cite{puzzle,petrarca}
$$BR_{SL}(B)|_{QCD}\geq 12.5\; \% \eqno(39)$$
It has been known for some time that the measured semileptonic
branching ratios fall below the predicted ones. Now, however,  one
has reached a stage where one has to take such a deficit
seriously since both the experimental and the theoretical analyses
have become rather mature.

\subsubsection{ Fabula Docet?}

An intriguing problem has arisen, which warrants serious
consideration: how can one find an additional non-leptonic
enhancement of at least 15 to 20 \% to satisfy the bound of
eq.(30)?

There are various scenarios for resolving this apparent puzzle:

\noindent (i) Improved data could move $BR_{SL}(B)$ above 12 \% .

\noindent (ii) There could be a `cocktail', i.e. a combination of
several smallish effects all working in the same direction: the
experimental number could inch up; higher order $perturbative$
and non-perturbative corrections could turn out to be somewhat
larger than estimated by us.

\noindent (iii) Non-perturbative corrections could
conceivably be dramatically larger than anticipated by us. This
certainly would require going beyond the `Standard Version' of OPE;
for it would presumably mean that even in beauty decays there are
numerically significant non-perturbative corrections that enter
through the coefficients in the OPE. In that scenario one would
probably obtain considerably larger differences in the lifetimes of
beauty hadrons than stated in eqs.(22, 25) above!

\noindent (iv) The most intriguing possibility would be the
intervention of New Physics in $B$ decays.

\subsubsection{Lepton Spectra}

The method outlined above has
been extended to treat the lepton energy spectra
in $H_b\ra l\nu X$ transitions.
The expansion is now in $1/(1-y)m_b$ rather
than in $1/m_b$, where $y=E_l/E_l^{max}$ denotes the normalized lepton energy
(for $b\ra u$ decays $y=2E_l/m_b\,$).
This expansion is obviously and
necessarily singular at $y=1$, i.e. in the
endpoint region, and one has to apply care in interpreting the results
there.

To order $1/m_b^2$ the
spectrum $d\Gamma /dy$ is evaluated without any free
parameters with the non-perturbative corrections
entering through the expectation values
$G_B$ and $K_B$; in practice there is at present some
numerical uncertainty in the
size of the kinetic energy term $K_B$, as already mentioned.
The shape of the spectrum thus derived from QCD
turns out to be remarkably
similar to the one
obtained from the phenomenological $AC^2M^2$ model~\cite{alta}
that has been fitted to the data. An important element of that
model was the introduction of a Fermi motion ascribed to the
heavy flavour quark. It has been found~\cite{bigi2,mn,motion}
that this
notion of Fermi motion finds a natural home also in a rigorous
QCD treatment: one can define a universal distribution function
that describes the motion of the beauty quark inside $H_b$
irrespective of the specifics of the decay process; to obtain the
observable spectrum predicted from QCD one has to
fold this distribution function with the primary lepton
spectrum from beauty decays
\footnote{One should remember however that the universality
holds only as long as the mass of the final state quark is the same. In
QCD the corresponding functions that enter here are quite different for
$b\ra c$ and for $b \ra u\,$ (or $\,b\ra s+\ga$) transitions.}.
The properties of this distribution function control the
shape of the lepton spectrum in the endpoint region.
The second moment of the distribution function is given by
$\aver{\vec p\, ^2}$. One can also show that this function
as a matter of
principle cannot be obtained from a non-relativistic
hadronic wavefunction, although the $b$ quark Fermi motion
is of a non-relativistic nature~\cite{motion}.

Lastly, one finds that sizeable differences can arise in the endpoint
spectrum of $B_d$ vs. $B^-$ mesons to order $1/m_b^3$ due to WA in the KM
suppressed decays~\cite{wa}. A detailed study of this difference in
the spectra can provide information about the four-fermion matrix
elements driving WA in nonleptonic decays and affecting PI.

\subsection{Other Inclusive Non-Leptonic Decays}

The $1/m_b$ expansion allows not only to calculate the overall
non-leptonic and semileptonic rates, but also various sub-classes,
like, e.g., those non-leptonic transitions that are driven
by $b\ra c\bar cs$, $b\ra u\bar ud$ etc. While we have expressed some
notes of caution about the prediction for $b\ra c\bar cs$ reactions since
the energy release there is not very large, no such caveat applies to the
$b\ra u\bar ud$ process: the expression in eq. (33) is easily adapted by
the obvious substitutions: $V_{ub}$ for $V_{bc}$ and $m_u=0$ for $m_c$;
the expectation values of the local operators remain the same. Whether
the prediction for the KM suppressed, inclusive non-leptonic $B$ decays
can ever be tested with any decent accuracy, is of course a different matter.

\subsection{Radiative Beauty Decays}

The non-perturbative contributions to
$\Gamma (H_b\ra s+\gamma )$ through order $1/m_b^2$
are obtained in a straightforward manner~\cite{manifesto}:
$$\Gamma (H_b\ra X_s+\gamma )=\Gamma (b\ra s\gamma )
\left( \frac{\matel{H_b}{\bar bb}{H_b}}{2M_{H_b}}-
\frac{\matel{H_b}{\bar bi\sigma \cdot Gb}{H_b}}{2M_{H_b}m_b^2}
+...
\right)
\eqno(40)$$
where we have set $m_s=0$. From eqs.(40) and
(32) one reads off that the ratio
$\Gamma (H_b\ra s+\gamma)/\Gamma (H_b\ra l\nu X)$ remains
practically unaffected by the non-perturbative corrections.

More interesting effects arise in the photon spectrum:
to lowest order it is given by a single line reflecting the
two-body nature of $b\ra s+\gamma$;
both perturbative and
non-perturbative corrections turn it into a continuous spectrum.
Yet this will not be discussed here; the reader is referred to a
quickly expanding literature~\cite{motion,neubert,wise}
on that topic.

\subsection{Charm Multiplicity in the Final State}

Because of the KM hierarchy $|V(cb)|^2\gg |V(ub)|^2$ and
since the
transition $b\ra c\bar cs$ occurs one realizes that
$$N_{charm}\equiv \frac{\mbox{Number of charm states}}
{B \mbox{ decay}} > 1\eqno(41)$$
has to hold where charmonia states enter into the book keeping with a charm
multiplicity of two. Since there is not enough phase-space for
$b\ra c\bar ud,\, cl\nu$ to transmogrify itself into
$B\ra D\bar DD,\, D\bar DDl \nu$ etc., one knows without ado
that each of these transitions will yield exactly one charm state
per $B$ decays. It used to be stated that one actually predicts
$N_{charm}=1.15$ because of $BR(b\ra c\bar cs)=0.15$. The prediction
for the latter
quantity depends of course quite sensitively on the values
adopted for the ratio of the charm to the beauty quark mass, as
expressed by the function $I_0(x,x,0)$ defined in eq. (35a)
with $x=(m_c/m_b)^2$. Using values for $m_c$ and $m_b$ as
inferred
from Heavy Quark Expansions one actually finds
$$N_{charm}\sim 1.2\div 1.3\eqno(42)$$
The data exhibit a considerably lower charm content,
namely~\cite{danilov}
$$N_{charm}=0.932\pm 0.10\; \; \; ARGUS\eqno(43a)$$
$$N_{charm}=1.026\pm 0.057\; \; \; CLEO\eqno(43b)$$
Taking eqs.(42, 43) at face value one would have to state the
existence of a significant `charm deficit'. On the other hand one
should keep in mind that in the transition $b\ra c\bar cs$ the
energy release is not that large. Thus the $1/m_b$ expansion has to
be applied with a grain of salt here: the non-perturbative
as well as perturbative
corrections in $b\ra c\bar cs$ could be larger than expected and
negative. But one conclusion can be drawn in any event:
attempting to lower the predicted semileptonic branching ratio for
$B$ mesons by increasing $BR(b\ra c\bar cs)$ would fail!

\subsection{Final State Interactions}

Invoking strong final state interactions (FSI) -- in the form of phase
shifts and absorption -- to escape conflict with the data represents a
time-honoured and popular tool in phenomenological analyses; FSI are typically
introduced in an ad-hoc fashion with an obscure dynamical foundation. Two
attitudes towards FSI seem to prevail in the community: (a) It constitutes
an {\em a priori} hopeless enterprise to account for FSI even in a
semi-quantitative
way because they are generated by
strong re-scattering processes among real hadrons.
(b) Whatever the origin of FSI, they are
actually suppressed in heavy flavour decays.

Such a situation is quite unsatisfactory, not because FSI are by themselves
enlightening -- they certainly are not -- but because they represent a
{\em conditio sine qua non} for the observability of direct CP violation in $B$
decays. Attitude (b) would suggest that direct CP asymmetries will be small in
$B$ decays; while attitude (a) on the other hand holds out the hope that one
might find sizeable
direct CP asymmetries, it would make their interpretation quite ambiguous.

The experience we have gained recently from the QCD treatments of heavy flavour
decays leads us to the following expectation: FSI in $exclusive$ decays
may well possess a rather complex and non-trivial pattern that we cannot
predict at present; yet re-scattering in $inclusive$ transitions must be
treatable
in QCD using the same methodology that has been introduced in the previous
sections.

One estimates FSI effects in non-leptonic $B\ra X\ind{charmless}$
decays by analyzing the interference
of the Penguin amplitude $b\ra s[d]\bar qq$ with the tree-level KM suppressed
$b\ra u\bar us[d]$ amplitude~\cite{soni}. These two amplitudes possess a
``hard'' FSI phase
difference $\delta _P$ due to the on-mass-shell intermediate $c$ and $\bar c$
quarks in the Penguin process; one easily finds $\delta _P\sim 0.1$. It was
shown~\cite{uri} that
all higher order gluon corrections can be summed up in a compact way; they
reduce $\delta _P$ by just $\sim $ 20 \%. This demonstrates
that the perturbative expansion
makes sense also for this quantity and thus refutes earlier claims to the
contrary.
An elegant analysis can be given~\cite{uri} that
involves rather general arguments based on gauge invariance and the equation of
motion in QCD, and shows the strong suppression of the decays of the type
$b\ra s + gg[ ggg]\,$; the latter can actually be traced back to the first
genuine QCD study of strange decays~\cite{penguin}.

Only the purely perturbative corrections to FSI have been addressed in
ref.~\cite{uri}.
Non-perturbative contributions still await a detailed analysis; on general
grounds one
expects them to be sizeable, yet at the same time theoretically tractable!

\subsection{$B$ Decays into Charmonia}
\nopagebreak

A very intriguing class of reactions is provided by the decays into
charmonia states: $B\ra \psi /\psi '/\chi +X$. No rigorous QCD
analysis of these modes has been given yet, but reasonable
phenomenological treatments do exist. Folding the wavefunctions for
colour singlet $\bar cc$ boundstates with the $\bar cc$ spectrum in
$b\ra c\bar c s$ one can make predictions about the relative as well as
the absolute rates~\cite{rueckl}:
$$\Gamma (b\ra \psi _{dir}X):\Gamma (b\ra \psi 'X):
\Gamma (b\ra \eta _cX):\Gamma (b\ra \chi _1X)
\simeq $$
$$\simeq
1:0.31:0.57:0.27\eqno(44a)$$
Furthermore
$$\Gamma (b\ra \chi _2X)=0\eqno(44b)$$
holds in this ansatz.
These ratios depend of course on the wave functions chosen, but are
independent of $c_1,\, c_2$ and $N_C$. The absolute widths on the
other hand depend on these parameters
$$\Gamma (b\ra \psi _{dir}X)\simeq (0.42\div 0.45)(c_2+
\frac{1}{N_C}c_1)^2\Gamma _0\eqno(45)$$
while being rather insensitive to the values
adopted for the
quark masses. For the part of the $\psi$ wave function that is
relevant here is probed also in $\psi \ra e^+e^-$ and can therefore
be obtained from the data on that electromagnetic decay.
In eq.(45) the width for $directly$ produced $\psi$ was given;
including the feed-down from $b\ra \psi '/\chi _1+X$ one
finds for the total rate for $\psi$ production:
$$\Gamma (b\ra \psi +X)\simeq (0.52\div 0.56)
(c_2+\frac{1}{N_C}c_1)^2\Gamma _0\eqno(46)$$
and, therefore,
$$BR(B\ra \psi +X)\simeq (0.27\div 0.4)\,\% .
\eqno(47a)$$
If the $1/N_C$ term in eq. (46) is omitted one gets, instead,
$$
BR(B\ra \psi +X)\simeq
(0.83\div 1.2)\, \% .
\eqno(47b)$$
Comparing these predictions with the data from ARGUS
and CLEO
$$BR(B\ra \psi X)=(1.09\pm 0.04\pm 0.07)\% \eqno(48a)$$
$$BR(B\ra \psi 'X)=(0.30\pm 0.05\pm 0.03)\% \eqno(48b)$$
$$BR(B\ra \chi _1X)=(0.54\pm 0.15\pm 0.14)\% \eqno(48c)$$
one draws three conclusions:

\noindent $\bullet$ The prediction on $BR(B\ra \psi 'X)$ vs.
$BR(B\ra \psi X)$ is in good agreement with the data.

\noindent $\bullet$ For the natural value $1/N_C=1/3$ the improved
parton model expression in eq.(46) yields a branching ratio that is
too low by a factor of $\sim 3-4$. With $1/N_C=0$ the observed
branching ratio is reproduced, as it is for $1/N_C\simeq 0.45$!
This means -- not surprisingly -- that for $b\ra c\bar cs$
transitions the deviations from this model
are large and not fully implemented through employing
a hadronic wavefunction. It remains a challenge to theory to
identify those corrections.

\noindent $\bullet$ The prediction on
$BR(B\ra \chi _1X)$ vs. $BR(B\ra \psi X)$ is
$\sim 2\; \sigma$ low. This could be blamed on a less than
optimal choice for the $\chi _1$ wavefunction. As already mentioned
the relevant part of the $\psi$ and $\psi '$ wavefunctions is
directly calibrated by the data on $\psi ,\, \psi '\ra
e^+e^-$; for the $\chi _1$ wavefunction such a cross check
does not (yet) exist and one has to rely on a specific ansatz
for the interquark potential, which introduces additional
systematic uncertainties. There exists of course another
possibility~\cite{braaten}:
the $c\bar c$ pair could be produced in
a colour $octet$ configuration which then
sheds its colour charge through emission
of a soft gluon and transforms itself into a colour singlet
P wave charmonium state. Naive factorization does not hold anymore
in the presence of this second source of $\chi$ production in
$B$ decays; its weight depends on the probability for
a colour octet $c\bar c$ to transmogrify itself into
a P-wave charmonium state which cannot be predicted from
first principles at present. On the other hand one can
extract its size from the requirement to saturate the
observed $B\ra \chi _1X$ rate; subsequently one can then predict
the rate for $B\ra \chi _2X$. One finds
$$\Gamma (B\ra \chi _1X)\sim \Gamma (B\ra \chi _2X)$$
in contrast to eq. (44).

\section{Exclusive Two-Body Decays of Beauty}

A discussion of exclusive non-leptonic decays has to be opened
with a note of caution:

{\em First Theoretical Caveat: The relationship between
inclusive and exclusive transition rates is
nothing short of delicate!}

This piece of common sense can be illustrated by the following
example~\cite{bigi0}. Consider the {\em corrections} to the decay
width of a
$(Q\bar q)$ meson that are induced by Weak Annihilation, see
Fig.~2

As indicated there are three cuts in the
$Q\bar q \ra Q\bar q$ forward scattering amplitude
representing different final states, one with an
on-shell gluon and the other two with a (slightly)
off-shell gluon in the form of a $\bar q q$ pair.
Summing over these three cuts yields an
overall correction $|T|^2$ that remains finite even in the limit
$m_q/m_ Q\ra 0$~\cite{bigi0,bigi3}.
However a very striking pattern emerges when one
considers separately the three `exclusive' channels $(a)$,
$(b)$ and $(c)$
\footnote{In the real world these three channels can of course
not be distinguished; yet this academic model can illustrate the
relevant point.}:
The contribution $(b)$ constituting the square of an
amplitude is positive; in the limit $m_q/m_Q\ra 0$ it is dominated
by a term $+(m_Q^2/m_q^2)|T|^2$ with $T$ denoting a
quantity that is regular in the limit $m_q/m_Q\ra 0$; the contributions $(a)$
and
$(c)$ on the other hand represent interference terms that,
taken together, are of
the form $-(m_Q^2/m_q^2)|T|^2$ for $m_q/m_Q\ra 0$ so that the
sum of $(a)$, $(b)$ and $(c)$
possesses a regular limit.
We want to draw the following lesson from this discussion: a
small effect in an overall rate can be due to large
cancellations among subclasses of decays. The relevance of this
statement will become clearer later on.

\subsection{Phenomenological Models}

A popular phenomenological model of non-leptonic two-body decays
of charm and beauty was suggested in
ref.~\cite{bsw}.
There are three main ingredients in all models of this type: \break
(i) One assumes
factorization, i.e. one uses $\matel {M_1M_2}{J_{\mu}J_{\mu}}
{D} \simeq \matel{M_1}{J_{\mu}}{D}\cdot \matel{M_2}{J_{\mu}}
{0}$ to describe $D\ra M_1M_2$. (ii) One employs one's
favourite hadronic wavefunctions to compute
$\matel{M_1}{J_{\mu}}{D}$. Very recently Heavy Quark Symmetry
and Chiral Symmetry (for the
light quarks) have been incorporated into these
wavefunctions~\cite{gatto}.
(iii) All two-body modes are then
expressed in terms of two free fit parameters $a_1^{(c)}$
and $a_2^{(c)}$,
with $a_1^{(c)}$ controlling the `class I'
$D^0\ra M_1^+M_2^-$
and $a_2^{(c)}$ the `class II'
$D^0\ra M_1^0M_2^0$ transitions;
both quantities contribute coherently to the `class III'
transitions $D^+\ra M_1^0M_2^+$. The analogous procedure is
applied to  $B$ decays allowing though for
$a_1^{(b)}\neq a_1^{(c)}$ and $a_2^{(b)}\neq a_2^{(c)}$.

With these two free parameters $a_{1,2}^{(c)}$ (and some
considerable degree of poetic license in invoking strong
final state interactions) one obtains a decent fit for the
$D^0$ and $D^+$ modes (much less so, however, for $D_s$
decays). The
situation is rather similar in $B$
decays.
One has to
point out, though, that this success is helped by a considerable
ambiguity in the estimates of the matrix elements of the
type $\langle \pi |J_\mu | B\rangle $ and
by the forgiving imprecision in many of the branching ratios
measured so far.

Exclusive decay rates depend sensitively on
long-distance dynamics. In the factorization approximation
large distances  enter  through simple hadronic
matrix elements $\matel {M}{J_{\mu }}{M}$ and
$\matel{M}{J_{\mu}}{0}$. The coefficients
$a_1$ and $a_2$ are:
$$a_1=c_1+\xi c_2\eqno(49a)$$
$$a_2=c_2+\xi c_1\eqno(49b)$$
where
$c_1$ and $c_2$, see eq. (17), include the radiative QCD corrections
due to hard gluons; the quantity $\xi$ is introduced as a fudge factor.
Literally speaking, factorization implies $\xi =1/N_C=1/3$. Quite
often $\xi$ is treated as a fit parameter
reflecting our ignorance of how a quark and an anti-quark
that are $not$ correlated in colour combine to form a meson.
Thus, deviations of $\xi$ from $1/3$ parametrize, in a certain way,
non-factorizable contributions. The non-factorizable contributions show
up even in perturbation theory, but these seem to be numerically small.
An example of the non-perturbative contribution which can violate
factorization is
provided by the strong
final state interactions (FSI). The latter can easily change
the value of $\xi$. Some prominent features
of FSI are actually added in an ad-hoc fashion; yet even so
one would a priori $not$ expect that all these
non-perturbative corrections can by and large be lumped into a
single fudge factor $\xi$.
At this point we would like to express another note of
caution:

{\em Second Theoretical Caveat: There exists no
general proof of the Dogma of Factorization for real hadrons; we
actually
consider it unlikely to be of universal validity,
in particular
for class II and III transitions.}
It thus makes eminent sense to subject this dogma to as many
different experimental tests as possible.

A fit to eleven exclusive hadronic $B$ branching ratios
yields~\cite{waldi}
$$a_1=1.05\pm 0.03\pm 0.10\eqno(50a)$$
$$a_2=0.227\pm 0.012\pm 0.022\eqno(50b)$$
A few comments are in order here:

(i) The results stated in eqs. (50) are bad news for
a popular  program to infer $a_1$ and $a_2$ from
perturbative QCD plus factorization:
the effective transition operators
are renormalized with coefficients $c_+,c_-$ generated
from perturbative QCD; then the  hadronic matrix elements
are factorized. The coefficient $a_2$ does
not come out correctly. Moreover, the so-called
rule of $1/N_C$ (this  is an {\em ad hoc} assumption that
only the terms leading in $1/N_C$ are to be retained),
which miraculously helped in the two-body decays of charm,
seems only to worsen the situation for beauty.
The  program
``factorization plus the rule of $1/N_C$" implies
$a_1=c_1,\; a_2=c_2$,
  in clear conflict with the findings
in $B$ decays, eqs. (50), since $c_2$ is negative!
Yet even before the following statement applied:

\noindent {\em Third Theoretical Caveat: It is very unlikely that the
rule of retaining only leading terms in $1/N_C$ is
universally
implemented in QCD.} Our analysis of inclusive heavy-flavour
decays actually found cases where this rule was dynamically
realized (a) for $D$ as well as for $B$ decays,
(b) for $D$, but not for $B$ decays, or (c) for
neither~\cite{bigi0,bigi1,BLOKI} (see also Sect. 4.2).

(ii) The most striking feature of eqs. (50) is that the relative
sign between $a_1$ and $a_2$ is positive, i.e. that a
$constructive$ interference occurs in class III transitions,
i.e. in exclusive two-body $B^-$ decays! This is surprising in
four aspects: (a) It is in clear contrast to the situation in $D$
decays where the $D^+$ modes suffer from a destructive interference.
(b) While it does not pose any fundamental problem for the
BSW model, it represents a basic failing for the $1/N_C$ ansatz
which predicts a destructive interference in the two-body
modes of $B$ as well as of $D$ decays.
(c) The observed
enhancement of the $B^+$ rates is remarkably uniform~\cite{waldi}:
$$BR(B^-\ra D^0\pi ^-)/BR(B^0\ra D^+\pi ^-)=1.71\pm 0.38
\eqno(51a)$$
$$BR(B^-\ra D^0\rho ^-)/BR(B^0\ra D^+\rho ^-)=1.60\pm 0.46
\eqno(51b)$$
$$BR(B^-\ra D^{*0}\pi ^-)/BR(B^0\ra D^{*+}\pi ^-)=1.79\pm 0.39
\eqno(51c)$$
(d) It raises the question of whether
the same constructive interference
might occur for the {\em inclusive} rate thus shortening
$\tau (B^-)$ relative to $\tau (B_d)$ rather than
lengthening it.

As stated in Sect. 3 the $B^-$ lifetime is predicted to
exceed the $B_d$ lifetime, albeit by a few percent only;
yet one has
to keep in mind the {\em First Theoretical Caveat} stated in the
beginning of this section: a small correction in the inclusive
rate is quite likely to be made up by large contributions
of alternating signs coming from different
classes of exclusive transitions.

For a better understanding of these problems one has to
progress to treatments that are rooted more firmly in QCD.
However we would like to first stress the important lessons
we have learnt and are still learning from these
phenomenological descriptions:

-- They have yielded quite a few successful `predictions' in a
user-friendly way providing a useful pattern for cataloguing dozens of
the decay rates, a reference frame for
more sophisticated approaches.

-- They have helped us considerably in focusing on the underlying
theoretical problems such as the question of factorization or the
$1/N_C$ rule.

-- From their fits to the data they provide us with valuable,
albeit indirect information on the final state interactions.
Such information is crucial in studies of direct CP violation.

\subsection{Theoretical Analysis}

The first treatment of two-body decays of
heavy flavour hadrons that is
intrinsically connected to QCD was given in ref.~\cite{BLOKI}
some time ago for $D$ decays, within the framework of the  QCD sum rules.
To extend this analysis to non-leptonic $B$ decays is a
rather non-trivial undertaking; work on
treating $B\ra J/\psi K$ through QCD is in progress now~\cite{RUECKLPRIVATE}.

Recently there has been progress in analysing $deviations$ from
the $1/N_C$ rule (see Sect. 4.1)  in some exclusive two-particle $B$
decays~\cite{BLOKII}. It has been shown how
non-perturbative effects in QCD can
provide a dynamical realization
of this rule  in some decay
channels, but not in others. The key role
is played by the same  chromomagnetic operator $\sigma \cdot G$ that
was repeatedly discussed above in connection with the inclusive decays.
It generates non-perturbative
corrections that are specific for the decay channel considered,
yet can be estimated in a model-independant way.

The general method can be illustrated through the example of the
decay $B^0\ra D^+\pi ^-$.
The non-factorizable terms in this decay are reducible to the
amplitude $\langle D\pi |{\cal L}_{oct} |B\rangle $
where
$${\cal L}_{oct}\sim (\bar c \gamma_\mu (1-\gamma_5)t^ab)
(\bar d\gamma_\mu (1-\gamma_5)t^a u) ,
$$
with $t^a$ denoting the colour matrices. Colour has to be exchanged
between the brackets because otherwise the light quarks can not form
the pion; this can be done by a soft gluon.

To estimate this effect
one considers the time ordered correlation function
$$ {\cal A}^{\beta }\equiv \int d^4xe^{iqx}
\matel{D}{T\{ {\cal L}_{oct}(x),A^{\beta}\} }{\bar B}\eqno(52)$$
where an auxiliary  axial current $A^{\beta}=\bar u\gamma^{\beta}
\gamma_5d$ annihilates the pion and $q$ is an external momentum
flowing through $A^{\beta}$.
To calculate the correlator one adopts a similar procedure as
in QCD sum rules. After continuing ${\cal A}^{\beta}$ into the
Euclidean region $-q^2=Q^2\sim 1\; \GeV ^2$ one invokes
duality in the following way: on the one hand one expresses
${\cal A}^{\beta}$ in terms of $M_{non-fact}$,
the non-factorizable part of the amplitude, as obtained from
${\cal L}_{oct}$:
$${\cal A}^{oct}(Q^2)=M_{non-fact}\frac{f_{\pi}q^{\beta}}{q^2}
+...\eqno(53)$$
where $+...$ denotes the contributions from the higher resonances
produced by the axial current $A^{\beta}$; on the other hand one
applies an OPE to ${\cal A}^{\beta}$ to find
$${\cal A}^{oct}(Q^2)=i\frac{1}{4\pi ^2}
\frac{q_{\alpha}q_{\beta}}{q^2}\matel{D}{\bar c\gamma _{\mu}\gamma_5
{\tilde G}_{\alpha \mu}b}{B}+...\eqno(54)$$
where $+...$ now denotes pre-asymptotic corrections from
higher dimensional operators. If both masses, $m_c$ and $m_b$, are
treated as heavy, the matrix element in eq. (54) is related to the expectation
value of the chromomagnetic operator $\sigma G$ and is expressed in
terms of $\mu_G^2$, see eq. (36b).
Comparing the leading terms in eqs. (53, 54) one obtains for
the ratio of the non-factorizable to the $1/N_C$
factorizable parts of the amplitude
$$r_{non-fact}=-\frac{N_C\mu _G^2}{4\pi ^2f_{\pi}^2}\eqno(55)$$
An  essential difference to the
usual QCD sum rules analysis is worth noting: rather than the vacuum
condensates we deal here with hadronic expectation
values between
heavy flavour hadron states. Numerically one finds
$$r_{non-fact}(B^0\ra D^+\pi ^-)\sim -1\eqno(56)$$
and likewise for the mode $B^0\ra D^+\rho ^-$; i.e. in these
two modes non-factorizable contributions which are necessarily
of order $1/N_C$ basically cancel against the
non-leading factorizable contributions of order $1/N_C$ thus
leading to a dynamical realization of the $1/N_C$ rule. The
findings are similar for $B\ra D\bar D$ decays. Yet
the situation changes for other modes:
in $B\ra D^*\bar D^*$ the non-factorizable
contributions are suppressed; likewise in
$B^0\ra D^{*+}\pi ^-,\;D^{*+}\rho ^-$.

There are two general conclusions we want to draw from this
analysis: (i) The weight of non-factorizable contributions is
indeed quite channel dependent -- as expected.
(ii) The discussion so far was given for class I transitions where
we found that the coefficient $a_1$ is not quite universal, but
has some channel dependence. The situation is much more complex and
actually at present unclear for class II (and III) transitions. For
the $a_2$ amplitude contains the matrix element
$\matel{B}{\bar b\gamma _{\nu}g\tilde G_{\alpha \nu}u}{\pi}$.
This formfactor, unfortunately,  can  not be determined from HQET.

This treatment has reached so far only the qualitative or at best
semi-quantitative stage. No detailed analysis has yet been given
about the question whether one can really suppress the contributions
from higher resonances and from higher dimensional operators in the
sum rule of eqs.(53, 54) to a sufficient degree. Furthermore,
the operator product expansion, eq. (54), assumes that
$(M_B-M_D)/(M_B+M_D)<<1$; in the real world this parameter is rather
of order unity.
 Also radiative corrections
have not been included yet. Further details can be found in ref.~\cite{BLOKII}.
Thus, much more theoretical work is needed before such a
treatment finds its definite form.

\subsection{Prizes to be Attained}

The theoretical methods one applies to exclusive decays are
often not of the most lucid kind. Yet they
are essential (if imperfect) tools for addressing fundamental
questions. Let us cite just one topical example:

$\Delta \Gamma (B_s)$, i.e. the lifetime
difference between $B_{s,short}$ and $B_{s,long}$,
is usually computed from the quark box diagram with internal
$c$ (and $u$) quarks, leading to a result like
the one quoted in eq. (15).
However the weight of such a short-distance contribution to
$\Delta \Gamma (B_s)$ is much more uncertain than that of the
local contribution to $\Delta m(B_s)$, which is given
by virtual top exchanges. It is therefore
conceivable in principle that nontrivial long-distance dynamics could
provide in reality a larger
contribution. This possibility has been analysed by the Orsay
group~\cite{ORSAY} considering separate
transitions
$$B_s\ra D_s^{(*)}\bar D_s^{(*)}, \psi \phi, \psi \eta
\ra \bar B_s\; $$
and it has been inferred from present data
$\Delta \Gamma/\bar{\Gamma} (B_s)\simeq 0.15$
in satisfactory agreement with what one finds using the quark box
diagram, eq. (24).

\section{Summary and Outlook}

\subsection{Status}
New and more powerful second-generation
theoretical technologies are
emerging: QCD sum rules, Heavy Quark Symmetry, $1/m_Q$
expansions and lattice simulations of QCD. They are leading
to

$\bullet$ significant conceptual progress, namely a better
understanding of (i) the form and size of non-perturbative
corrections,
(ii) the relationship between charm and
beauty decays, where the former play the role of a microscope
for the non-perturbative corrections in the latter,
and (iii) the differences and similarities of baryon vs. meson
decays;

$\bullet$ the realization that charm and beauty baryons deserve
detailed studies in their own right.

$\bullet$ a quantitative phenomenology that is genuinely based
on QCD:
$$\frac {\tau (D^+)}{\tau (D^0)}\sim 2; \; \;
BR_{SL}(D^+)\sim 16\%,\; \; BR_{SL}(D^0)\sim 8\%;
$$
$$\frac{\tau (D_s)}{\tau (D^0)}\sim 1.0\pm \mbox{ a few per
cent}\; ,\eqno(57)$$
i.e. the data are reproduced within the accuracy of the
expansion.
$$\frac{\tau (B^-)}{\tau (B_d)}\simeq
1+0.05\cdot \frac{f_B^2}{(200\; \MeV )^2}; \; \; \;
BR_{SL}(B)\geq
12\%
\eqno(58)$$
$$\frac{\Delta \Gamma (B_s)}{\bar \Gamma (B_s)}
\simeq 0.18\cdot \frac{f_{B_s}^2}{(200\; \MeV )^2},\;\; \; \;
\frac{\tau (\Lambda _b)}{\tau (B_d)}
\sim 0.9
\eqno(59)$$
$$d\Gamma (B^-\ra l\nu X)/dE_l\neq d\Gamma (B_d\ra l\nu X)/dE_l
\eqno(60)$$
The prediction for $BR_{SL}(B)$ is somewhat larger than
present CLEO and ARGUS measurements. This could turn
out to be a
serious -- or intriguing -- discrepancy. It could
conceivably signal the presence of anomalously
large higher-order contributions that so far have not
been included in the
theoretical expression. In that case one would expect
lifetime ratios for beauty hadrons to differ more from
unity than stated in eqs. (58) and (59)~\cite{puzzle}.

 Phenomenological models for nonleptonic two-body modes
are encountering discrepancies with more precise
data; yet they
continue
to be useful and help us in focusing on the underlying theoretical
issues.

\subsection{Future}
One can expect a refinement of and
increased cooperation (rather than just coexistence) between
the second-generation theoretical technologies.
On the experimental side one can hope for

-- lifetime measurements for individual beauty hadrons with
$10\%$ accuracy soon and percent accuracy in the longer run;

-- data on $\tau (B_s)$ {\em separately} from
$B_s\ra \psi \phi$ and from $B_s\ra l\nu D_s$;

-- perform the `class I, II, III' phenomenology
{\em individually} for {\it KM} allowed and {\it KM}
suppressed $B$  decays.

The primary goal in all these efforts is to be able
to exploit to the fullest over
the next 20 years or so, the
discovery potential or even discovery guarantee
that awaits us in beauty physics. It certainly would be
a crime not to make all conceivable efforts to obtain the
required experimental facilities.

\vspace*{0.5cm}

{\bf Acknowledgements:} \hspace{.4em}
This work was supported by the National
Science Foundation under grant number PHY 92-13313 and
by DOE under the grant number DOE-AC02-83ER40105.
Two of us (I.B. and N.U.) gratefully acknowledge
computing support from
the Wisconsin-ALEPH group in preparing this text.

\newpage
\vspace*{2cm}
\hspace*{3em}{\Large{\bf FIGURE CAPTIONS}}
\vspace*{.6cm}

{\bf Fig. 1} \\
Corrections to the $b\bar q \ra b\bar q$
forward scattering amplitude induced by Weak Annihilation.
The three cuts (a), (b) and (c) represent different
final states for the $b\bar q$ decay.
\vspace*{.3cm}

{\bf Fig. 2}\\
{\bf (a)} Diagram generating the operator $\bar bb$.\\
{\bf (b)} Diagram describing Weak Annihilation.\\
{\bf (c)} Diagram describing Pauli Interference.


\begin{thebibliography}{99}

\bibitem{wilson}
K. Wilson, {\it Phys. Rev.} {\bf 179} (1969) 1499;\\
K. Wilson and J. Kogut, {\it Phys. Reports} {\bf 12} (1974) 75.

\bibitem{bilic}
N. Bili\'{c}, B. Guberina, J. Trampeti\'{c}, {\it Nucl. Phys.}
{\bf B248} (1984) 261;\\
M. Voloshin, M. Shifman, {it Yad. Fiz.} {\bf 41} (1985) 187
[{\it Sov. Journ.
Nucl. Phys.} {\bf 41} (1985) 120]; {\it ZhETF} {\bf 91} (1986) 1180
[{\it Sov. Phys. -- JETP} {\bf 64} (1986) 698].

\bibitem{chay}
J. Chay, H. Georgi, B. Grinstein, {\it Phys. Lett.} {\bf B247} (1990)
399.

\bibitem{bigi0}
I.I. Bigi, N.G. Uraltsev, {\it Phys. Lett.} {\bf 280B} (1992) 120.

\bibitem{bigi1}
I. Bigi, N. Uraltsev, A. Vainshtein, {\it Phys. Lett.} {\bf B293}
(1992)
430; (E) B297 (1993) 477;\\
B. Blok, M. Shifman, {\it Nucl. Phys.} {\bf B399} (1993) 441; 459;
\\
I. Bigi, B. Blok, M. Shifman, N. Uraltsev, A. Vainshtein, {\it Proc. of
the
1992 DPF meeting of APS}, Fermilab, November 1992  [Preprint
UND-HEP-92-BIG07].

\bibitem{bigi2}
I. Bigi, M. Shifman, N. Uraltsev, A. Vainshtein, {\it Phys. Rev.
Lett.} {\bf 71} (1993) 496.

\bibitem{HQS}
N. Isgur, M. Wise, {\it Phys. Lett.} {\bf B232} (1989) 113; {\bf B237}
(1990) 527;\\
see also
M. Voloshin, M. Shifman, {\it Yad. Fiz.} {\bf 47} (1988) 801
[{\it Sov. J. Nucl. Phys.} {\bf 47} (1988) 511].

\bibitem{EHQT}
E. Eichten, B. Hill, {\it Phys. Lett.} {\bf B234} (1990) 511;\\
H. Georgi, {\it Phys. Lett.} {\bf B240} (1990) 447.

\bibitem{EHQTr}
For recent reviews see\\
H. Georgi, in {\it Proceedings of TASI-91}, eds. R. K. Ellis et al., (World
Scientific, 1991);\\
B. Grinstein,  in {\it High Energy Phenomenology}, eds. R. Huerta and
M.A. P\'{e}res (World Scientific, 1991);\\
N. Isgur, M. Wise, in {\em Heavy Flavours}, eds. A.J. Buras and M.
Linder (World Scientific, 1992);\\
T. Mannel, in {\it QCD - 20 Years Later}, eds. P. M. Zerwas and H.A.
Kastrup (World Scientific, 1993);\\
M. Neubert,  Preprint SLAC-PUB-6263, 1993 [{\it Phys. Reports, to be
published}].

\bibitem{SV1}
M. Shifman, M. Voloshin, 1982, see in
V. Khoze, M. Shifman, {\it Uspekhi Fiz. Nauk} {\bf 140} (1983) 3
[{\it Sov. Phys.
Uspekhi}
{\bf 26} (1983) 387].

\bibitem{bigi3}
I. Bigi, N. Uraltsev, Preprint CERN-TH.7020/93, to appear
in {\it Nucl. Phys.} {\bf B}.

\bibitem{buvpre}
I. Bigi, N. Uraltsev, A. Vainshtein, Preprint FERMILAB-PUB-92/158-T;
this formula was dropped for the article that appeared in Phys.Lett.B, first
ref.~\cite{bigi1}, in order not to exceed the maximum length.

\bibitem{motion}
I. Bigi, M. Shifman, N. Uraltsev, A. Vainshtein, preprint
CERN-TH.7129/93, 1993.

\bibitem{braun}
P.Ball, V.Braun, Munich preprint MPI-PH-93-51; see also:
M.Neubert, {\em Phys.Rev.} {\bf D46} (1992) 1076; for a
criticism see: M. Neubert, preprint CERN-TH.7070/93.

\bibitem{them}
A.V.Manohar, M.B.Wise, San Diego preprint UCSD/PTH 93-14,
hep-ph/9308246.

\bibitem{alta}
G.Altarelli et al., {\em Nucl.Phys.} {\bf B208} (1982) 365.

\bibitem{mn}
M. Neubert, Preprint CERN-TH.7087/93.

\bibitem{pich}
E. Braaten, S Narison, A. Pich, {\it Nucl. Phys.}
{\bf B373} (1992) 581;
F. Le Diberder, A. Pich, {\it Phys. Lett.} {\bf B286}
(1992) 147; {\it Phys. Lett.} {\bf B289} (1992) 165.

\bibitem{hybrid}
M. Voloshin and M. Shifman, {\it Sov. J. Nucl. Phys.}
{\bf 45} (1987) 292;\\
H.D. Politzer and M. Wise, {\it Phys. Lett.} {\bf B206} (1988) 681.

\bibitem{hybrid1}
M. Shifman, M. Voloshin, {\it Sov.Phys. - JETP} {\bf 64} (1986)
698.

\bibitem{pi}
B. Guberina et al., {\it Phys. Lett.} {\bf B89} (1981) 303.

\bibitem{bloks}
B. Blok, M. Shifman, {\em Talk at the Third Workshop on the
Tau-Charm Factory}, June 1993, Spain [Preprint TPI-MINN-93/55-T,
to be published in the Proceedings].

\bibitem{bsosc}
M.Voloshin, M. Shifman, N.Uraltsev, V.Khoze,
{\it Sov.J.Nucl.Phys.} {\bf 46} (1987) 112.

\bibitem{sharma}
V. Sharma, F. Weber, this book.

\bibitem{argus}
H. Albrecht et al. (ARGUS Collab.), DESY Preprint DESY 93-104, 1993.

\bibitem{cleo}
J. Bartelt et al. (CLEO Collab.), Cornell University Preprint
CLEO-CONF 93-19, 1993.

\bibitem{puzzle}
I. Bigi, B. Blok, M. Shifman, A. Vainshtein, Preprint
CERN-TH.7082/93.

\bibitem{petrarca}
G. Altarelli, S. Petrarca, {\it Phys. Lett.} {\bf B261}
(1991) 303.

\bibitem{manifesto}
I. Bigi, B. Blok, M. Shifman, N. Uraltsev, A. Vainshtein,
Preprint UND-HEP-92-BIG07, in: Proceedings of the APS-DPF'92
Meeting, Fermilab, November 1992.

\bibitem{soni}
M.Bander, D.Silverman, A.Soni, {\it Phys. Rev. Lett.}
{\bf 43} (1979) 242.

\bibitem{uri}
Yu.L.Dokshitzer, N.G.Uraltsev, {\it JETP Lett.} {\bf 52} (1990) 509;\\
N. Uraltsev,
Preprint UND-HEP-92-BIG08, in: Proceedings of the APS-DPF'92
Meeting, Fermilab, November 1992.

\bibitem{penguin}
A. Vainshtein, V.Zakharov, M. Shifman, {\it Sov.Phys. JETP} {\bf 45} (1977)
670.

\bibitem{neubert}
M. Neubert, Preprint CERN-TH. 7113/93.

\bibitem{wise}
A. Falk et al., Preprint UCSD/PTH 93-38.

\bibitem{danilov}
M. Danilov, Rapporteur Talk at the 1993 EPS Conference,
Marseille, 1993.

\bibitem{rueckl}
J. H. K\"{u}hn, R. R\"{u}ckl, {\it Phys. Lett.} {\bf 135B} (1984)
477; {\it ibid.} {\bf B258} (1991) 499.

\bibitem{braaten}
G. Bodwin, E. Braaten, T.C. Yuan, P. Lepage,
{\it Phys.Rev.} {\bf D46} (1992) R3703.


\bibitem {wa}
I.I.Bigi, N.G.Uraltsev, {\bf preprint} CERN-TH-7039/93 (1993);


\bibitem{bsw}
M.Bauer, B.Stech, {\it Phys.Lett.} {\bf B152} (1985)380;
M.Bauer, B.Stech, M.Wirbel, {\it Z.Phys.} {\bf C34}
(1987)103. For a more recent review, see:
M. Neubert, V. Rieckert, B. Stech, in: `Heavy Flavours',
ed. by A. Buras and M. Lindner, p. 286, World Scientific, 1992.

\bibitem{gatto}
A.Deandrea et al., {\it Phys. Lett.} {\bf B 318} (1993) 549.

\bibitem{waldi}
K. Honscheid, K. Schubert, R. Waldi, Preprint
OHSTPY-HEP-E-93-017.


\bibitem{wa}
I.I.Bigi, N.G.Uraltsev, preprint CERN-TH-7020/93 (1993).

\bibitem{BLOKI}
B.Blok, M.Shifman, {\em Sov.J.Nucl.Phys.} {\bf 45} (1987)
35,301,522.

\bibitem{RUECKLPRIVATE}
S. Belyaev. A. Khodjamirian and R. R\"{u}ckl, to be published.

\bibitem{BLOKII}
B.Blok, M.Shifman, {\em Nucl.Phys.} {\bf B389} (1993)534.


\bibitem{ORSAY}
R.Aleksan et al., Orsay preprint LPTHE 93/18.

\end{thebibliography}
\end{document}